\documentstyle[epsfig,aps]{revtex}

\newcommand{\ccap}[2]{\caption[#1]{#2}\label{#1}\vspace{0.2cm}}
\begin{document}
\draft
\title{ Anomalous WW$\gamma$ Vertex at LC+HERA Based 
$\gamma p$ Collider}
\mediumtext
\author{S. Ata\u{g} and \.{I}.T. \c{C}ak\i r} 
\address{Ankara University Department of Physics \\ Faculty of 
Sciences, 06100 Tandogan, Ankara, Turkey}
\maketitle

\begin{abstract}
The potential of LC+HERA based $\gamma$p collider to probe 
WW$\gamma$ vertex is presented through the discussion of 
sensitivity to anomalous couplings and $p_{T}$ distribution
of the final quark. The limits of  $-0.16<\Delta\kappa<0.14, 
\;\;\; -0.30<\lambda<0.30, \;\;\ $ at 95\% C.L. can be reached 
with integrated luminosity 100$pb^{-1}$ and they are 
competitive to  projected future limits from other colliders. 
The results are compared with corresponding ep collider using 
Weizs\"{a}cker-Williams Approximation.  
\end{abstract}

\vskip 0.5cm

\section{Introduction}
Recently there have been intensive studies to test the 
deviations from the Standard Model (SM) at present and 
future colliders. The investigation of three gauge boson 
couplings plays an important role to manifest the non 
abelian gauge symmetry in standard electroweak theory.
The precision measurement of the triple vector boson 
vertices will be the crucial test of the structure of the 
SM. 

Analysis of $WW\gamma$ vertex has been given by several 
papers for Fermilab\cite{abe}, LEP2\cite{bilenky} 
and ep\cite{dubinin,baur1,kim,baur2,janssen,noyes} 
collider DESY HERA. 
One of  further projects is the Linear Collider (LC)
 design at DESY \cite{trines} which is the only one
 that can be converted into an ep \cite{brinkmann,roeck} collider.
 An aditional advantage of the linac-ring type ep
 colliders is the possibility of building $\gamma$p
 colliders on their bases \cite{ciftci}. 
The high energy gamma beam
 is obtained by the Compton backscattering of laser
 photons off linac's electron beam. Estimations show
 that this possibility appears quite realistic, since
 the spectrum of the scattered laser photons is hard
 and the luminosity for $\gamma$p turns out to be of
 the same order as the one for ep collision.
According to present project at DESY 500 GeV electrons
are allowed to collide 820 GeV protons \cite{brinkmann,roeck}. 
In this case the corresponding parameters 
of the $\gamma$p collider
are shown in Table \ref{tab1}\cite{ciftci}.
For photoproduction processes the cross sections in
$\gamma$p machines are about one order of magnitude
larger than corresponding ep machines. Therefore such
kind of high energy $\gamma$p colliders will be
complementary tools to linac ring type ep colliders.

In this paper we examine the potential of future
LC+HERA based $\gamma$p collider to probe 
anomalous WW$\gamma$ coupling and compare the results
from ep collider. 

\begin{table}[bth]
\caption{Main parameters of LC+HERA based $\gamma$p
collider.
\label{tab1}}
\begin{center}
\begin{tabular}{lcccc}
Machine &$\sqrt{s_{ep}}$ TeV & $\sqrt{s_{\gamma p}}$ TeV
&$L_{\gamma p}(cm^{-1}s^{-1})$  \\
\hline
LC+HERA  &1.28 &1.16 & 2.5$\times 10^{31}$ \\
\end{tabular}
\end{center}
\end{table}

\section{ Lagrangian and Cross Sections}

C and P parity conserving effective lagrangian for two charged
and one photon interaction can be written following the papers
\cite{gaemers,hagiwara}

\begin{eqnarray}
L&&=e(W_{\mu\nu}^{\dagger}W^{\mu}A^{\nu}-
W^{\mu\nu}W_{\mu}^{\dagger}A_{\nu}+
\kappa W_{\mu}^{\dagger}W_{\nu}A^{\mu\nu} 
+{\lambda\over M_{W}^{2}}W_{\rho\mu}^{\dagger}
W_{\nu}^{\mu}A^{\mu\rho})
\end{eqnarray}
where

\begin{eqnarray}
W_{\mu\nu}=\partial_{\mu}W_{\nu}-\partial_{\nu}W_{\mu}\nonumber 
\end{eqnarray}
and dimensionless parameters $\kappa$ and $\lambda$ are related
to the magnetic and electric dipole moments.
$\kappa=1$ and $\lambda=0$ correspond Standard Model values.
In momentum space this has the following form with momenta
$W^{+}(p_1)$,$W^{-}(p_2)$ and $A(p_3)$

\begin{eqnarray}
\Gamma_{\mu\nu\rho}(p_{1},p_{2},p_{3})=&&e[g_{\mu\nu}
(p_{1}-p_{2}-{\lambda\over M_{W}^{2}}[(p_{2}.p_{3})p_{1}
-(p_{1}.p_{3})p_{2}])_{\rho}\nonumber \\
&&+g_{\mu\rho}(\kappa p_{3}-p_{1}
+ {\lambda\over M_{W}^{2}}[(p_{2}.p_{3})p_{1}
-(p_{1}.p_{2})p_{3}])_{\nu} \nonumber \\
&&+g_{\nu\rho}(p_{2}-\kappa p_{3}
-{\lambda\over M_{W}^{2}}[(p_{1}.p_{3})p_{2}
-(p_{1}.p_{2})p_{3}])_{\mu} \nonumber \\
&&+{\lambda\over M_{W}^{2}}(p_{2\mu}p_{3\nu}p_{1\rho}
-p_{3\mu}p_{1\nu}p_{2\rho}])]
\end{eqnarray}
where  $p_1+p_2+p_3=0$.

There are three Feynman diagrams for the subprocess
$\gamma q_{i}\rightarrow Wq_{j} $ and  only t-channel W
exchange graph contributes $WW\gamma$ vertex. Therefore
$\gamma p$ collision isolates $WW\gamma$ coupling but
corresponding ep process has the mixtures of $WW\gamma$
and WWZ couplings.

The unpolarized differential cross section for the subprocess
$\gamma q_{i}\rightarrow Wq_{j} $ can be obtained using
helicity amplitudes from \cite{baur2} summing over the helicities

\begin{eqnarray}
{d\hat{\sigma}\over d\hat{t}}={2\over{\hat{s}-M_{W}^{2}}}
{\beta\over {64\pi\hat{s}}} \sum_{\lambda_{\gamma}
\lambda_{W}}{1\over 2} M_{\lambda_{\gamma}\lambda_{W}}^{2}
\end{eqnarray}
where
\begin{eqnarray}
 M_{\lambda_{\gamma}\lambda_{W}}={e^{2}\over{\sqrt{2}
 \sin\theta_{W}}}{\hat{s}\over{\hat{s}+M_{W}^{2}}}
\sqrt{\beta}A_{\lambda_{\gamma}\lambda_{W}}\; , \;\; 
\beta=1-{M_{W}^{2}\over\hat{s}} 
\end{eqnarray}
and $\theta_{W}$ is the Weinberg angle. 

For the signal we are considering a quark jet and on-shell
W with leptonic decay mode
\begin{eqnarray}
\gamma p\rightarrow W+jet\rightarrow\mu+p_{T}^{miss}+jet
\end{eqnarray}
In this mode charged lepton and the quark jet are in general well
separated and the signal is in principle free of background of SM.

The total cross section for the subprocess 
$ \gamma q_{i} \rightarrow W q_{j}$ can be obtained as follows 
\cite{kim}:
\begin{eqnarray}
\hat{\sigma}=&&{{\alpha G_{F}M_{W}^{2}}\over{\sqrt{2}\hat{s}}}
|V_{q_{i}q_{j}}|^{2}\{(|e_{q}|-1)^{2}(1-2\hat{z}+2\hat{z}^{2})
\log({\hat{s}-M_{W}^{2}\over\Lambda^{2}})
-[(1-2\hat{z}+2\hat{z}^{2}) \nonumber \\
&&-2|e_{q}|(1+\kappa+2\hat{z}^{2})
+{{(1-\kappa)^{2}}\over{4\hat{z}}}-{{(1+\kappa)^{2}}\over{4}}]
\log{\hat{z}}+ [(2\kappa+{{(1-\kappa)^{2}}\over{16}})
{1\over \hat{z}} \nonumber \\
&&+({1\over 2}+{{3(1+|e_{q}|^{2})}\over{2}})\hat{z}
+(1+\kappa)|e_{q}|-{{(1-\kappa)^{2}}\over{16}}
+{|e_{q}|^{2}\over 2}](1-\hat{z}) \nonumber \\
&&-{{\lambda^{2}}\over{4\hat{z}^{2}}}(\hat{z}^{2}-2\hat{z}
\log{\hat{z}}-1)+{{\lambda}\over{16\hat{z}}}
(2\kappa+\lambda-2)[(\hat{z}-1)(\hat{z}-9)
+4(\hat{z}+1)\log{\hat{z}}]\}
\end{eqnarray}
where $\hat{z}=M_{W}^{2}/\hat{s}$ and $\Lambda^{2}$ is
cut off scale in order to regularize $\hat{t}$-pole of the
colinear singularity for massles quarks. In the case of massive
quarks there is no need such a kind of  cut. To get numerical
results we take into account the integrated cross section over
quark distributions inside the proton and the spectrum of the
backscattered laser photons\cite{ciftci}.

\begin{eqnarray}
\sigma=\int_{\tau_{min}}^{0.83}d\tau \int_{\tau/0.83}^{1}
{dx\over x}f_{\gamma/e}(\tau/x)f_{q/p}(x,Q^{2})\hat{\sigma}(\tau s)
\end{eqnarray}
with $\tau_{min}=(M_{W}+M_{q})^{2}/s$

In Table \ref{tab2} integrated total cross sections times branching
ratio of $W\rightarrow \mu\nu $ and corresponding number of events 
per year are shown for various values of
$\kappa$  and $\lambda$. Number of events has been calculated using 
\begin{eqnarray}
N=\sigma(\gamma p\rightarrow W+Jet)B(W\rightarrow \mu\nu)
A L_{int}
\end{eqnarray}
where  we take the acceptance in the muon channel A  and integrated 
luminosity $L_{int}$ as  65\% and 100pb$^{-1}$.
To give an idea about the comparison with corresponding ep
collider the cross sections obtained using Weizs\"{a}cker-Williams
approximation are also shown on the same table. Through
the calculations proton structure functions of Martin, Robert and
Stirling (MRS A)\cite{martin}  have been used with $Q^{2}=M_{W}^{2}$.
As the cross section $\sigma(\gamma p\rightarrow W+Jet)$ we have
considered the sum of $\sigma(\gamma u\rightarrow W^{+}d)$,
$\sigma(\gamma \bar{d}\rightarrow W^{+}\bar{u})$,
$\sigma(\gamma \bar{s}\rightarrow W^{+}\bar{c})$,
$\sigma(\gamma u\rightarrow W^{+}s)$,
$\sigma(\gamma \bar{s}\rightarrow W^{+}\bar{u})$, and
$\sigma(\gamma \bar{d}\rightarrow W^{+}\bar{c})$.

\begin{table}
\caption{Integrated total cross section times branching
ratio $\sigma(\gamma p\rightarrow Wj)\times B(W^{+}\rightarrow
 \mu\nu)$ in pb and corresponding number of events(in parentheses) 
for some $\kappa$ and $\lambda$ values.
\label{tab2}}
\begin{center}
\begin{tabular}{llcc}
Backscattered Laser  &  WWA & $\kappa$  & $\lambda $ \\
\hline
13.8(845) &1.3(85) &1 & 0 \\
25.1(1631) &1.9(123) &1 & 1 \\
59.0(3835) &3.7(240) &1 & 2 \\
9.7(631) & 0.9(60) &0 & 0 \\
23.4(1521) &2.1(137) &2 & 0 
\end{tabular}
\end{center}
\end{table}

As shown from Table \ref{tab2} the cross sections using backscattered
laser photons are considerably larger than the case of corresponding
ep collision. We assume that the form factor structure of
$\kappa-1$ and $\lambda$ do not depend on the momentum transfers
at the energy region considered.
Then anomalous terms containing
$\kappa$ grow with $\sqrt{\hat{s}}/M_{W}$ and terms with $\lambda$
rise with $\hat{s}/M_{W}^{2}$. 
Deviation $\Delta\kappa=\kappa-1=1$  from SM value changes the total
cross sections 30-70\% whereas the $\Delta\lambda=\lambda=1$ gives
80\% changes. Therefore high energy will improve the sensitivity
to anomalous couplings. For comparison with HERA energy
$\sqrt{s}=314$ GeV the similar results would be 20-40\% for
$\Delta\kappa=1$ and 5\% for $\Delta\lambda=1$.

It is important to see how the anomalous couplings change the 
shape of the transverse momentum distributions of the final quark 
jet. For this reason we  use  the following formula:

\begin{eqnarray}
{d\sigma\over dp_{T}}=&&2p_{T}\int_{y^{-}}^{y^{+}}dy
\int_{x_{a}^{min}}^{0.83}dx_{a} f_{\gamma/e}(x_{a})
f_{q/p}(x_{b},Q^{2}) 
({x_{a}x_{b}s \over {x_{a}s-2m_{T}E_{p}e^{y}}})
{d\hat{\sigma}\over d\hat{t}} 
\end{eqnarray}
where
\begin{eqnarray}
y^{\mp}=\log{[{{0.83s+m_{q}^{2}-M_{W}^{2}}
\over {4m_{T}E_{p}}}\mp\{ ({{0.83s+m_{q}^{2}-M_{W}^{2}}
\over {4m_{T}E_{p}}})^{2}-{0.83E_{e}\over E_{p}}\}^{1/2}]}
\end{eqnarray}
\begin{eqnarray}
&&x_{a}^{(1)}={{2m_{T}E_{p}e^{y}-m_{q}^{2}+M_{W}^{2}}
\over{s-2m_{T}E_{e}e^{-y}}} \;\; , \;\;\;\;\;
x_{a}^{(2)}={(M_{W}+m_{q})^{2}\over s} \nonumber \\
&&x_{a}^{min}=MAX(x_{a}^{(1)},x_{a}^{(2)}) \;\; , \;\;\;\;\;
x_{b}={{2m_{T}E_{e}x_{a}e^{-y}-m_{q}^{2}+M_{W}^{2}}
\over {x_{a}s-2m_{T}E_{p}e^{y}}}
\end{eqnarray}
with
\begin{eqnarray}
&&\hat{s}=x_{a}x_{b} \;\; , \;\;\;
\hat{t}=m_{q}^{2}-2E_{e}x_{a}m_{T}e^{-y} \;\; , \;\;\;
\hat{u}=m_{q}^{2}+M_{W}^{2}-\hat{s}-\hat{t} \nonumber \\
&&m_{T}^{2}=m_{q}^{2}+p_{T}^{2}
\end{eqnarray}
The $p_{T}$ spectrum B($W\rightarrow \mu\nu )
\times d\sigma/dp_{T}$ of the quark jet is 
shown in Fig. \ref{fig1} for various $\kappa$ and 
$\lambda$ values at LC+HERA based $\gamma$p collider.
Similar distribution is given in 
Fig. \ref{fig2} for Weizs\"{a}cker-Williams Approximation that 
covers the major contribution from ep collision.
It is clear that the cross section at 
large $p_{T}$ is quite sensitive to
anomalous $WW\gamma$ couplings. As $\lambda$ increases the cross 
section grows more rapidly when compared with $\kappa$ 
dependence at high $p_{T}$ region $p_{T}>100$ GeV. 
The cross sections with real gamma beam are one
order of magnitude larger than the case of WWA. 
Comparison between two figures also shows that the curves  
become more separable as $\hat{s}$ gets large.     

\newpage

\begin{figure}[htb]
  \begin{center}
  \epsfig{file=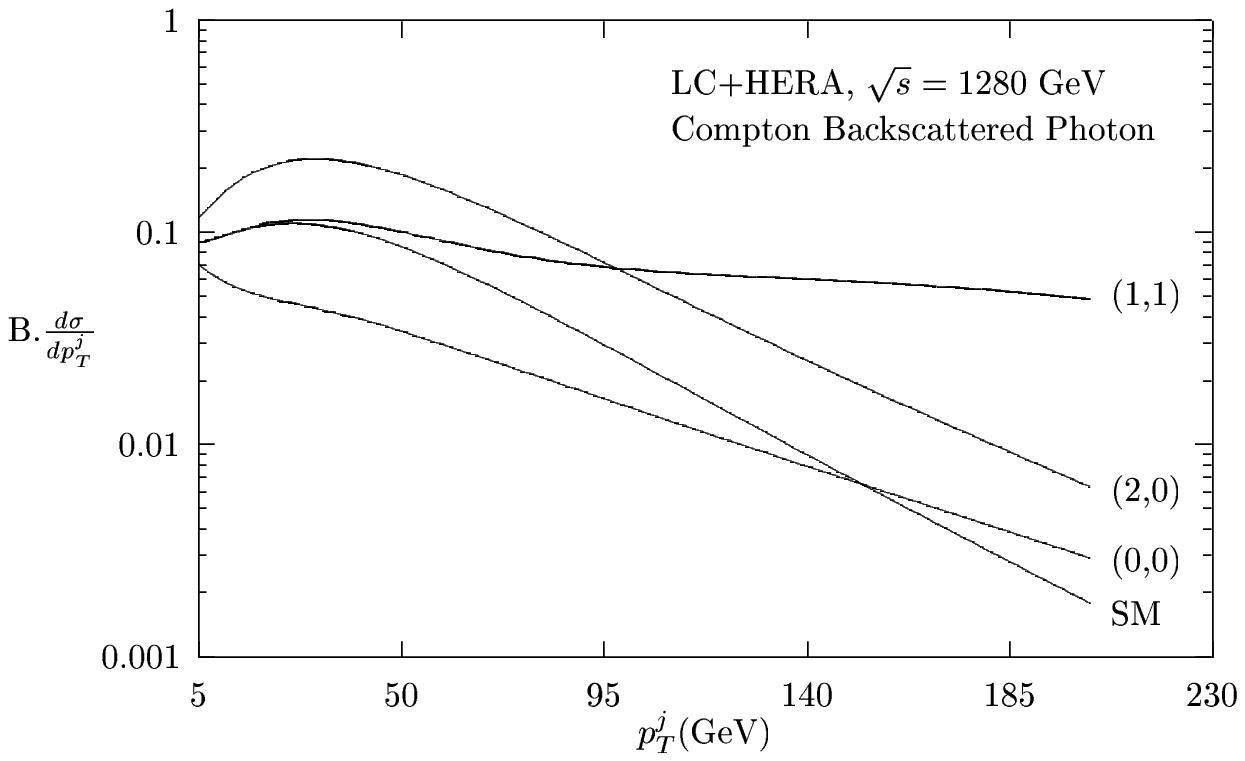}
  \end{center}
  \ccap{fig1}{\footnotesize  $\kappa$ and $\lambda$ dependence 
of the transverse momentum distribution of the quark jet at 
LC+HERA based $\gamma$p collider (Compton Backscattered Photon).
The unit of the vertical axis is pb/GeV and the numbers
in the parentheses stand for anomalous coupling parameters 
$(\kappa ,\lambda )$.  }
\end{figure}

\begin{figure}[htb]
  \begin{center}
  \epsfig{file=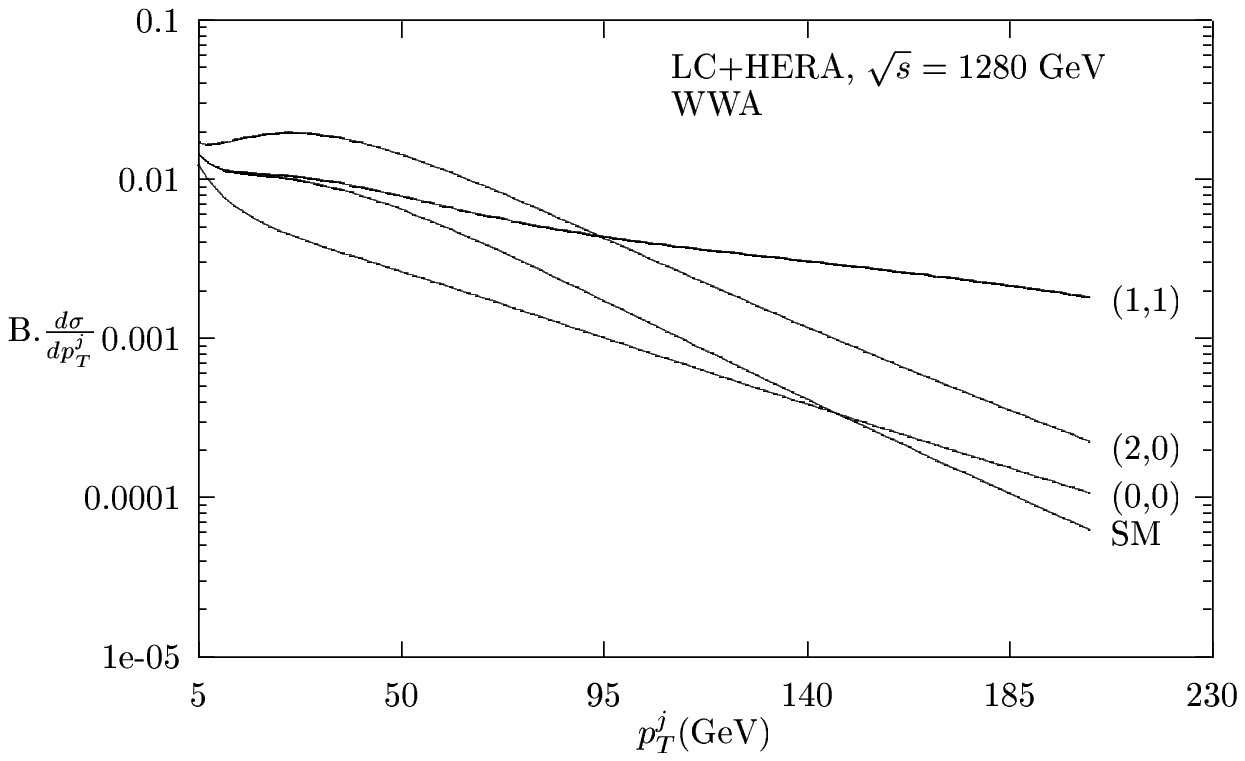}
  \end{center}
  \ccap{fig2}{\footnotesize  The same as the Fig.1 but for 
Weizs\"{a}cker Williams Approximation.}
\end{figure}

\section{Sensitivity to Anomalous Couplings}
   
We can estimate sensitivity of LC+HERA based $\gamma$p collider to
anomalous couplings by assuming the uncertainty in the luminosity 
measurement to be 2\% and systematic errors in the acceptance 
to be 2\% for $\mu\nu$ decay channel of the W and for the 
integrated luminosity values of $10^{2}$ pb$^{-1}$.
The acceptance in the muon channel is taken to be 65\%. 
We use the following expression  to add 
statistical and systematic errors to get 
the uncertainty on the cross section measurement
\begin{eqnarray}
{\Delta\sigma\over \sigma}=({1\over N} + ({\Delta L\over L})^{2}+
({\Delta A\over A})^{2})^{1\over 2}
\end{eqnarray}
where N is the number of $W^{+}$ and $W^{-}$ events given with 
their branching ratios in the $\mu\nu$ channel  
and A is the acceptance of the related channel. 
Above formula gives the following limits on the $\Delta\kappa$
and $\lambda$ for the deviation of the total cross section from 
the Standard Model value at 68\% and 95\% confidence levels:

\begin{center}
$-0.08<\Delta\kappa<0.08, \;\;\;\;  -0.22<\lambda<0.22, 
\;\;\;\;  $68\% C.L.\\
$-0.16<\Delta\kappa<0.14, \;\;\;\;  -0.30<\lambda<0.30,
\;\;\;\;   $ 95\% C.L.
\end{center}

On the ground of comparison we give the limits  at 
ep collider using Weizs\"{a}cker-Williams Approximation:

\begin{center}
$-0.20<\Delta\kappa<0.19, \;\;\;\;  -0.43<\lambda<0.43,
\;\;\;\;  $68\% C.L.\\
$-0.46<\Delta\kappa<0.37, \;\;\;\;  -0.62<\lambda<0.62,
\;\;\;\;   $ 95\% C.L.
\end{center}
In the Compton backscattered photon case the sensitivity 
is limited by the systematic errors and comparable to the limits 
from projected future colliders\cite{noyes}. However this is limited 
by the statistics  rather than systematic errors for WWA case.
Taking into account both the transverse momentum distribution of the 
quark jet and the sensitivity in the total cross section leads to 
very promising results for probing $WW\gamma$ vertex at $\gamma$p 
collider based on LC+HERA. 


\acknowledgements
Authors are grateful to the members of the AUHEP group 
and S. Sultansoy for drawing our attention to anomalous 
couplings.


\begin{thebibliography}{99}
\bibitem{abe} CDF Collaboration,F. Abe {\it et al.}, 
Phys. Rev. Lett.{\bf 74}, 1936 (1995); D0 Collaboration, 
Adachi {\it et al.} {\it ibid.} {\bf 75}, 1034 (1995).
\bibitem{bilenky} M. Bilenky, J.L. Kneur, F.M. Renard 
and Schildknecht, Nucl. Phys. {\bf B409}, 22 (1993);
F.A. Berends and A. van Sighem, Nucl. Phys. {\bf B454},
467 (1995); C.G. Papadopulos, Phys. Lett {\bf B352}, 144 
(1995); F. Boudjema, in Proceedings of the Workshop Physics
and Experiments with Linear Colliders, edited by A. Miyamoto,
Y. Fuji, T. Matsui and S. Iwata (World Scientific, Singapore,
1996) Vol. I, p. 199.
\bibitem{dubinin} M.N. Dubinin and H,S. Song, Phys. Rev. 
{\bf D57}, 2927 (1998).
\bibitem{baur1} U. Baur, J. Vermaseren and D. Zeppenfeld,
Nucl. Phys. {\bf B375}, 3 (1992)
\bibitem{kim} C.S. Kim, Jungil Lee and H.G. Song, 
Z. Phys. {\bf C63}, 673 (1994).
\bibitem{baur2} U. Baur and D. Zeppenfeld, Nucl. 
Phys. {\bf B325}, 253 (1989).
\bibitem{janssen} M. Janssen, Z. Phys. {\bf C52}, 165, 1991;
M. B\"{o}hm and A. Rosado, {\it ibid.} {\bf 39}, 275 (1988).
\bibitem{noyes}V.A. Noyes, Proceedings of the Workshop 
on Future Physics at HERA 1995/96, edited by Ingelman, 
a. De Roeck and R. Klanner, p. 190.
\bibitem{trines} D. Trines, Proceedings of the International 
Workshop on the Linac-Ring Type ep and $\gamma$p Colliders, 
published in Turkish J. Phys. {\bf 22}, 529 (1998).
\bibitem{brinkmann} R. Brinkmann, {\it ibid.}, {\bf 22}, 661 (1998); 
Z.Z. Aydin, A.K. Ciftci and S. Sultansoy, Nucl. Instrum. 
and Meth. {\bf A351}, 261 (1994).
\bibitem{roeck} A. De Roeck, Turkish J. Phys. {\bf 22}, 595 (1998).
\bibitem{ciftci} S.I. Alekhin {\it et al.}, 
Int. J. Mod. Phys. {\bf A6}, 21 (1991);
A.K. Ciftci, S. Sultansoy, S. Turkoz and O. Yavas,
Nucl. Instrum. and Meth. {\bf A365}, 317 (1995); Z.Z. Aydin 
{\it et al.}, Int. J. Mod. Phys. {\bf A11}, 2019 (1996)
A.K. Ciftci, Turkish J. Phys. {\bf 22}, 675 (1998).
\bibitem{gaemers} K.J.F. Gaemers and G.J. Gournaris, Z. Phys. 
{\bf C1}, 259 (1979).
\bibitem{hagiwara} K. Hagiwara, R.D. Peccei, D. Zeppenfeld and 
K. Hikasa, Nucl. Phys. {\bf B282}, 253 (1987).
\bibitem{martin}A.D. Martin, W.J. Stirling and R.G. Roberts, 
Phys. Rev. {\bf D51}, 4756 (1995).
\end{thebibliography}
\end{document}